\documentclass[aps,twocolumn,prabib,amsmath,amssymb,superscriptaddress,notitlepage]{revtex4-1}
\usepackage{graphics}
\usepackage{bm}
\usepackage{dcolumn}
\usepackage{natbib}
\usepackage{epstopdf}
\usepackage{float}
\usepackage{graphicx}
\usepackage{epsfig}
\usepackage[pdfstartview=FitH]{hyperref}
\usepackage{color}
\usepackage{appendix}
\usepackage{ulem}

\newcommand{\rr}{{\bf r}}

\begin{document}
\title{Continuous phase transition between bosonic integer quantum Hall liquid and a trivial insulator:
Evidence for deconfined quantum criticality}
\author{Tian-Sheng Zeng}
\affiliation{School of Science, Westlake University, Hangzhou 310024, China and \\
Institute of Natural Sciences, Westlake Institute for Advanced Study, Hangzhou 310024, China}
\author{D. N. Sheng}
\affiliation{Department of Physics and Astronomy, California State University, Northridge, California 91330, USA}
\author{W. Zhu}
\affiliation{School of Science, Westlake University, Hangzhou 310024, China and \\
Institute of Natural Sciences, Westlake Institute for Advanced Study, Hangzhou 310024, China}

\date{\today}
\begin{abstract}

The deconfined quantum critical point, a prototype Landau-forbidden transition,
could exist in principle in the phase transitions involving a symmetry protected topological phase, however, examples
of such kinds of transitions in physical systems are rare beyond one-dimensional systems.
Here, using density-matrix renormalization group calculation,
we unveil a bosonic integer quantum Hall phase in two-dimensional correlated honeycomb lattice,
by full identification of its internal structure from the topological $\mathbf{K}$ matrix.
Moreover we demonstrate that imbalanced periodic chemical potentials can destroy the bosonic integer quantum Hall state
and drive it into a featureless trivial (Mott) insulator, where all physical observables evolve smoothly across the critical point.
At the critical point the entanglement entropy reveals a characteristic scaling behavior,
which is consistent with the critical field theory as an emergent QED$_3$ with two flavors of Dirac fermions.
\end{abstract}

\maketitle
\clearpage

\section{introduction}
Landau's spontaneously symmetry breaking mechanism and phase transition theory are cornerstones of modern condensed matter~\cite{Landau1937}. Following his steps, many research studies have focused on characterizing emerging quantum phases and their transitions.
An innovatory chapter developed in the past decade is topological phases of matter beyond the Landau's paradigm,
such as the symmetry protected topological (SPT) phases~\cite{Wen2017,Senthil2015}.
SPT phases are bulk gapped states with either gapless or degenerate edge excitations protected by specific symmetries,
and the prototype examples include topological insulators \cite{Kane2005}, Haldane spin chain~\cite{Haldane1983,Chen2013}, and
bosonic integer quantum Hall (BIQH) state~\cite{Senthil2013,Furukawa2013,Regnault2013,YHWu2013,Geraedts2013,He2015,Sterdyniak2015,Zeng2016},
just to name a few \cite{QiXL2011,Lu2012,Chen2012,Ashvin2013,Xu2013,Schuch2011}.
Despite the established SPT phases,
much less is understood about the quantum phase transitions between them.
It is speculated that the Landau-forbidden transitions can occur between distinct SPT phases,
or between a SPT phase neighboring with other trivial phases,
however, such exotic proposals are less explored due to limited examples beyond one-dimensional systems.   

\begin{figure}[b]
  \includegraphics[height=0.8in,width=2.9in]{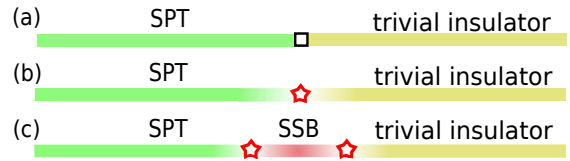}
  \caption{Possible phase transitions between a SPT phase and a trivial insulator:
  (a) first-order transition with a transition point (open square),
  (b) continuous phase transition with a quantum critical point (red star),
  and (c) an intermediate spontaneously symmetry-breaking (SSB) phase sandwiched in between.
  }\label{fig:spt}
\end{figure}

Generally, there exist three possible scenarios for phase transitions separating two distinct SPT phases,
or a SPT phase with another trivial phase as illustrated in Fig.~\ref{fig:spt}:
(i) a first-order transition, (ii) a direct continuous phase transition with a quantum critical point (QCP),
and (iii) a two-step transition via an intermediate symmetry-breaking phase.
There have been various works to characterize the critical theory between one-dimensional SPT phases.
Compared to first-order transitions identified in models with discrete symmetries ~\cite{Gu2009,Pollmann2010,ZXLiu2011} [Fig.~\ref{fig:spt}(a)],
a second-order phase transition between SPT phases occurs [Fig.~\ref{fig:spt}(b)],
if in the presence of a continuous symmetry~\cite{Nijs1989} or
fractionalized excitations corresponding to delocalized boundary excitations of the non-trivial SPT phase~\cite{Chen2013b,Tsui2017,Ardonne2015,Ruben2017}.
However the above knowledge is obtained in one dimension,
and much less is known in higher dimensions than one, which raises a fundamental question:
can a continuous quantum phase transition between two
SPT phases happen in higher dimensions?

Indeed, recent two-dimensional QCPs and their related theories for  Landau-forbidden transitions  have received  particular interests.
A special QCP known as the deconfined quantum critical point,
provides an opportunity to study the properties of deconfined excitations and
emergent gauge fields~\cite{Senthil2004a,Senthil2004b} as such a QCP controls the scaling behavior of
physical quantities.
Recent discovery of the connection between symmetry and duality~\cite{Karch2016,CWang2017,CWang2018}
has stimulated additional numerical studies of this exotic scaling behavior at this QCP~\cite{YQQin2017,Scott2017}.
Moreover, it has been proposed that a class of Landau-forbidden transitions with deconfined quantum criticality occurs
between bosonic SPT phases in two dimensions~\cite{Grover2013,Lu2014}, such as two-flavor fermionic massless quantum electrodynamics (QED) in (2+1) dimensions. In a web of duality, this fermionic $N_f=2$ QED$_3$ is dual to the bosonic easy-plane noncompact CP$^1$ (NCCP$^1$) model which describes a putative continuous phase transition between the N\'{e}el and valence bond solid phases in two-dimensional quantum magnets~\cite{CWang2017,YQQin2017}. Such a dual description opens up a new way to understand the phase transition between different symmetry-broken phases with distinct Landau order parameters.
However, numerical Monte Carlo simulations on a designed lattice model with $U(1)\times U(1)$ symmetry
gives a weakly first-order transition from BIQH to trivial insulator~\cite{Scott2017},
at odds with the continuous scenario.
For another two-dimensional model with discrete $Z_2$ symmetry,
only a first-order transition was observed~\cite{Morampudi2014}.
Taken as a whole, in space dimension higher than one,
it remains unclear whether a microscopic model can exhibit
a continuous phase transition between a SPT phase and a trivial insulator, which is one of the key focuses in our work.

This paper is organized as follows. In Sec.~\ref{model}, we introduce the interacting Hamiltonian of hardcore bosons loaded on a correlated honeycomb lattice at half filling, and describe the general phase diagram under the interplay of periodic potential and nearest-neighboring interaction. In Sec.~\ref{biqh}, we numerically demonstrate the emergence of the bosonic integer quantum Hall effect, based on topological information of the $\mathbf{K}$ matrix,
including quantized drag Hall conductance.
In Sec.~\ref{criticality}, we discuss the continuous phase transition between BIQH liquid and a trivial (Mott) insulator induced by the periodic potential,
and identify the entanglement signature at this critical point. Finally, in Sec.~\ref{summary}, we summarize our results and discuss the prospect of investigating the phase transition between BIQH liquid and a trivial insulator.

\section{Models and Methods}\label{model}

\begin{figure}[t]
  \includegraphics[height=1.61in,width=1.68in]{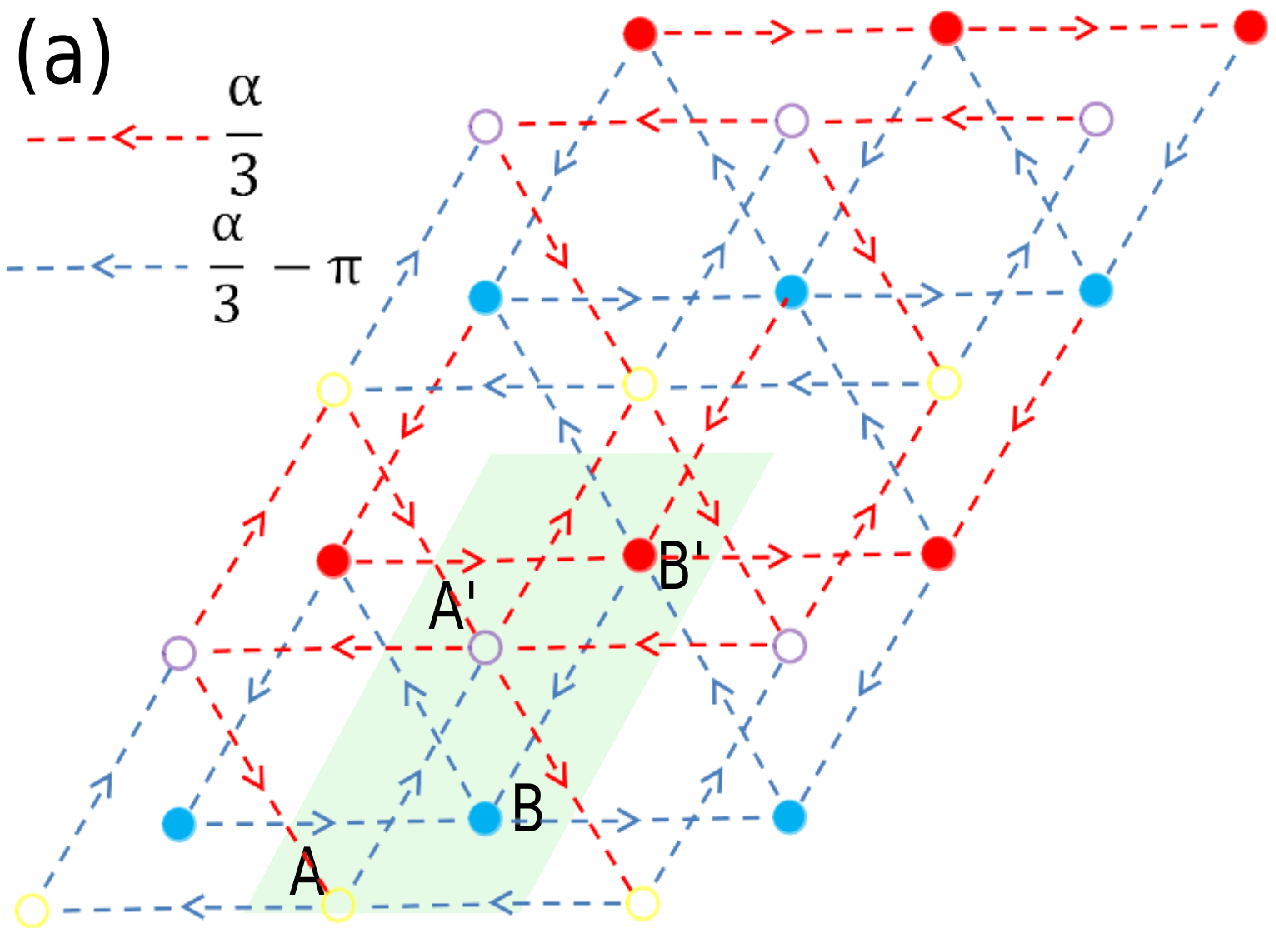}
  \includegraphics[height=1.66in,width=1.68in]{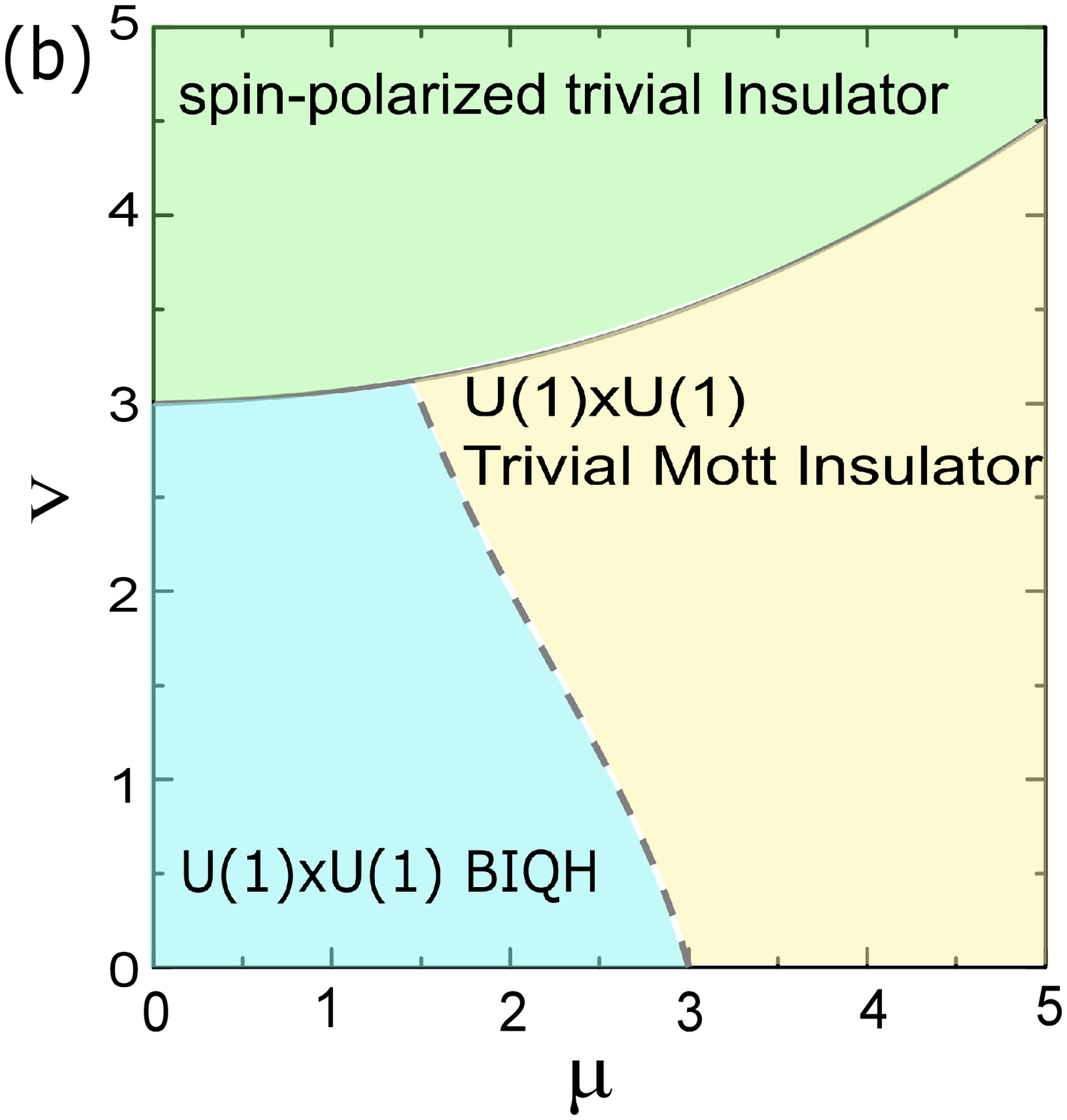}
  \caption{(a) The schematic plot of the correlated honeycomb lattice model with chiral background flux.
  Sublattice $A$ ($B$) is labeled by open (filled) circles.
  The arrow link shows the correlated hopping direction carrying chiral flux $\phi=\alpha/3$, $\alpha/3-\pi$ (here we take $\alpha=\pi/2$),
  therefore each triangular plaquette forms a flux $\pi-\alpha$ or $\alpha$ and each hexagonal plaquette has a flux $2\alpha$. The unite cell contains four inequivalent sites $A,A',B,B'$ in the shaded region.
  (b) The global quantum phase diagram. The phase transition between BIQH and trivial insulator (marked by dashed line) is continuous, where the phase boundary is determined by the singular peak of the correlation length in DMRG, while the solid line depicts the first-order transition, where the phase boundary is determined by the discontinuity of the first-order derivative of the ground state energy. }\label{lattice}
\end{figure}

In this work, we consider a topological honeycomb lattice filled with interacting hardcore bosons at half filling, similar to that in Refs.~\cite{He2015,Fuji2016}.
The bond hopping between sublattice $A$ ($B$) sites is coupled to the particle number of the nearest-neighbor $B$ ($A$) site:
\begin{align}
&H=\sum_{\langle \langle ij \rangle\rangle} \left[e^{i\phi_{ij}} (2 n_k^b-1) a^\dag_i a_j +h.c.\right]+\sum_{j}\mu_j(n_{j}^{a}+n_{j}^{b}) \nonumber\\
&+\sum_{\langle \langle kl \rangle\rangle} \left[e^{i\phi_{kl}}(2 n_j^a-1) b^\dag_k b_l + h.c.\right] + V\sum_{\langle ik \rangle} n_{i}^{a}n_{k}^{b},\label{ham}
\end{align}
where $a$ ($b$) is the particle annihilation operator on sublattice $A$ ($B$),
and $n_{i}^{a} (n_{k}^{b})$ is the particle number operator on sublattice $A$ ($B$) of the honeycomb lattice [see Fig.~\ref{lattice}(a)]. Here the chiral phases $\phi_{ij}=\alpha/3$ or $(\alpha/3-\pi)$ are taken according to the colored arrow links, and the basic unit cell contains four inequivalent sites $A,A',B,B'$ as a result of $\phi_{ij}$  as indicated in Fig.~\ref{lattice}(a).
$\mu_j$ is the periodic potential with commensurate period two: $\mu_j=\mu$ for $A,B$ sites
while $\mu_j=0$ for $A',B'$ sites within each unit cell. $V$ denotes the interaction strength on the nearest-neighbor pair of sites. Below we regard sublattice $A$ ($B$) as the spin degree of freedom $\uparrow$ ($\downarrow$), and take Eq.~\ref{ham} as an effective two-component lattice model.

Our main findings are summarized in the phase diagram in Fig.~\ref{lattice}(b).
In the parameter region $V<V_c$ and $\mu<\mu_c$,
we find a robust BIQH phase. The BIQH
phase is characterized by a single gapped ground state
on torus geometry, and its topological nature will be
identified by the quantized Chern number
and the topological $\mathbf{K}$ matrix as elaborated later.
We also show that the BIQH liquid phase is neighboring with
two types of topologically trivial insulators.
While the interaction $V$ drives a first-order phase transition,
the ground state undergoes a continuous phase transition from BIQH to a featureless insulator induced by periodic potential $\mu$.
Remarkably, we find a qualitative evidence for the critical theory as QED$_3$ with two flavors of Dirac fermions.

In order to study the ground state
and related phase transitions, we implement the density-matrix renormalization group (DMRG) algorithm \cite{white1992}
combined with exact diagonalization (ED) calculations,
both of which have been proven to be powerful tools
for lattice models containing arbitrary strong and frustrated interactions.
In the ED study, we explore the many-body ground state of $H$ in a finite system
with the total number of sites $N_s=L_x\times L_y$.
The ED calculations on the periodic lattice are limited to smaller systems,
while  we exploit infinite DMRG~\cite{Ian2008,Cincio2013} for cylinders with infinite length $L_x$ and finite width $L_y$ (we use even $L_y$
between $6$ and $12$, and periodic boundary conditions in the $y$ direction).
We keep the number of states up to $M=6500$ to obtain accurate results for the different system sizes we studied.

\section{BIQH phase}\label{biqh}

Now we begin with a discussion of the topological nature of the $\mathbf{K}$ matrix of a two-component BIQH phase.
For generic two-component systems at a given filling,
they can be classified by a class of the integer-valued symmetric matrix $\mathbf{K}=\begin{pmatrix}
m & n\\
n & m'\\
\end{pmatrix}$.
The $\mathbf{K}$ matrix is related to the Hall conductance (the Chern number matrix $\mathbf{C}$ for a two-component system), through
$\mathbf{C}=\mathbf{K}^{-1}=\begin{pmatrix}
C_{\uparrow\uparrow} & C_{\uparrow\downarrow} \\
C_{\downarrow\uparrow} & C_{\downarrow\downarrow} \\
\end{pmatrix}$.
Here the off-diagonal part $C_{\uparrow\downarrow}$ is related to the drag Hall conductance.

\begin{figure}[t]
  \includegraphics[height=2.1in,width=3.2in]{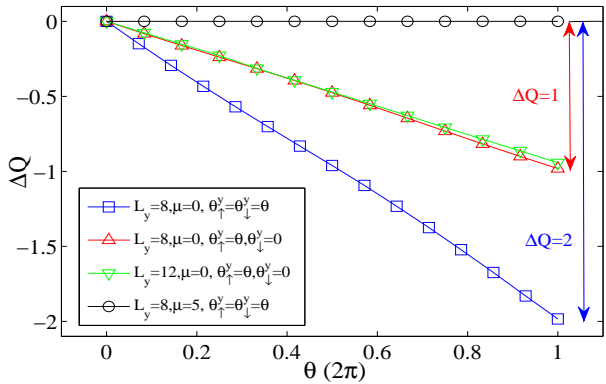}
  \caption{\label{pump}(Color online)
  The charge transfer in the $x$ direction for hardcore bosons on the correlated honeycomb lattice
  with different cylinder widths at half filling under the insertion of two types of flux
  quantum along the $y$ direction: $\theta_{\uparrow}^{y}=\theta_{\downarrow}^{y}=\theta$ and $\theta_{\uparrow}^{y}=\theta,\theta_{\downarrow}^{y}=0$. Here the interaction $V=0$.}
\end{figure}

The Chern number matrix can be accessed in two independent and complementary ways.
The first way is to apply twisted boundary conditions
$\psi(\cdots,\rr_{\sigma}^{i}+L_{\alpha},\cdots)=\psi(\cdots,\rr_{\sigma}^{i},\cdots)\exp(i\theta_{\sigma}^{\alpha})$
where $\theta_{\sigma}^{\alpha}$ is the twisted angle for particles of the $\sigma$-th component in the $\alpha$ direction.
The many-body Chern number $C_{\sigma,\sigma'}$ of the ground state wavefunction $\psi(\theta_{\sigma}^{x},\theta_{\sigma'}^{y})$ is obtained by ~\cite{Sheng2003,Sheng2006}
\begin{align}
  C_{\sigma,\sigma'}=\int\frac{d\theta_{\sigma}^{x}d\theta_{\sigma'}^{y}}{2\pi} \mathbf{Im}\left(\langle{\frac{\partial\psi}{\partial\theta_{\sigma}^x}}|{\frac{\partial\psi}{\partial\theta_{\sigma'}^y}}\rangle
-\langle{\frac{\partial\psi}{\partial\theta_{\sigma'}^y}}|{\frac{\partial\psi}{\partial\theta_{\sigma}^x}}\rangle\right).\nonumber
\end{align}
For small system sizes $N_s=24,32$,
our ED study confirms the robustness of single ground state degeneracy. According to the field theory description of BIQH~\cite{Senthil2013,Lu2012}, this unique ground state is expected to host
\begin{align}\label{chern}
\mathbf{C}=\mathbf{K}^{-1}=\begin{pmatrix}
0 & 1\\
1 & 0\\
\end{pmatrix}.
\end{align}

\begin{figure}[t]
  \includegraphics[height=2.64in,width=3.4in]{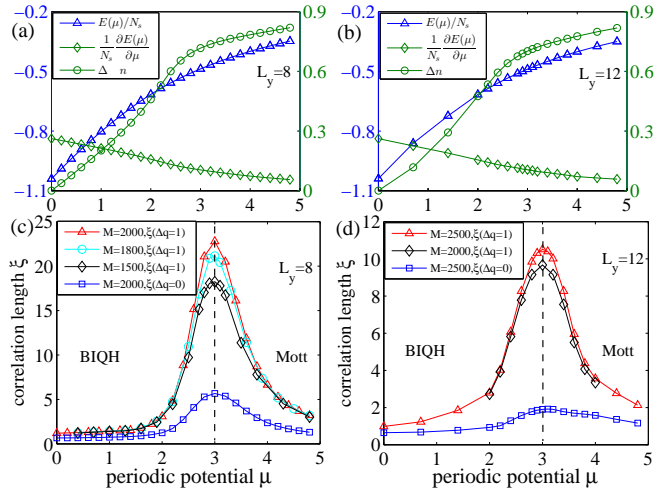}
  \caption{Numerical DMRG results on the second-order phase transition induced by periodic potential $\mu$ from BIQH phase to trivial (Mott) insulator at $V=0$.
  Top panels show the ground state energy per site, the first-order derivative, and density imbalance as a function of $\mu$.
  Bottom panels show the correlation lengths $\xi$ for different bond dimensions as a function of $\mu$.
  Left (right) panels are cylinder widths $L_y=8$ ($L_y=12$).
  The black dashed line locates the critical point where the peak position of correlation length is singular.
  }\label{fig:corr}
\end{figure}

Alternatively, we can approach the Chern number matrix in DMRG calculation.
We calculate the charge pumping of the ground state under the insertion of flux quantum on large cylinder systems,
in connection to the quantized Hall conductance~\cite{Gong2014}.
In DMRG we partition the cylinder along the $x$ direction into two halves.
By inserting one flux quantum $\theta_{\uparrow}^{y}=\theta,\theta_{\downarrow}^{y}=0$
from $\theta=0$ to $\theta=2\pi$ on the cylinder system,
the expectation value of the total particle number $\widehat{N}_{L}$ on the left side is
$N_{L}(\theta)=tr[\widehat{\rho}_L(\theta)\widehat{N}_{L}]$,
where $\widehat{\rho}_L$ is the reduced density matrix of the corresponding left part.
The net charge transfer from the right side to the left side of the system during each cycle is encoded by~\cite{Zeng2017,Zeng2018}
\begin{align}
  \Delta Q=N_{L}(2\pi)-N_{L}(0)=C_{\uparrow\uparrow}+C_{\downarrow\uparrow}.
\end{align}
Similarly, by inserting one flux quantum $\theta_{\uparrow}^{y}=\theta_{\downarrow}^{y}=\theta$ from $\theta=0$ to $\theta=2\pi$,
we obtain the charge pumping $\Delta Q=\sum_{\sigma,\sigma'}C_{\sigma,\sigma'}$. As shown in Fig.~\ref{pump}, for bosons at half filling,
by threading one flux quantum $\theta_{\uparrow}=\theta_{\downarrow}=\theta$ in both components, the total charge pump is just given by $\Delta Q=2$,
while a unit charge $\Delta Q\simeq1$ is pumped by threading one flux quantum $\theta_{\uparrow}^y=\theta,\theta_{\downarrow}^y=0$ in one component.
In light of these observations, we determine the Chern number matrix as Eq.~\ref{chern}.
Therefore, we establish that the unique ground state for bosons at half filling
is indeed a BIQH phase on a lattice system.
By contrast, when periodic potential is increased up to large values $\mu\gg1$,
we find that the expectation value of charge pumping $\Delta Q$ vanishes to a precision of $10^{-4}$,
signalling a topologically trivial (Mott) insulator phase with particles strongly localized at the lattice sites $j$ where $\mu_j=0$.

\begin{figure}[t]
  \includegraphics[height=1.75in,width=1.50in]{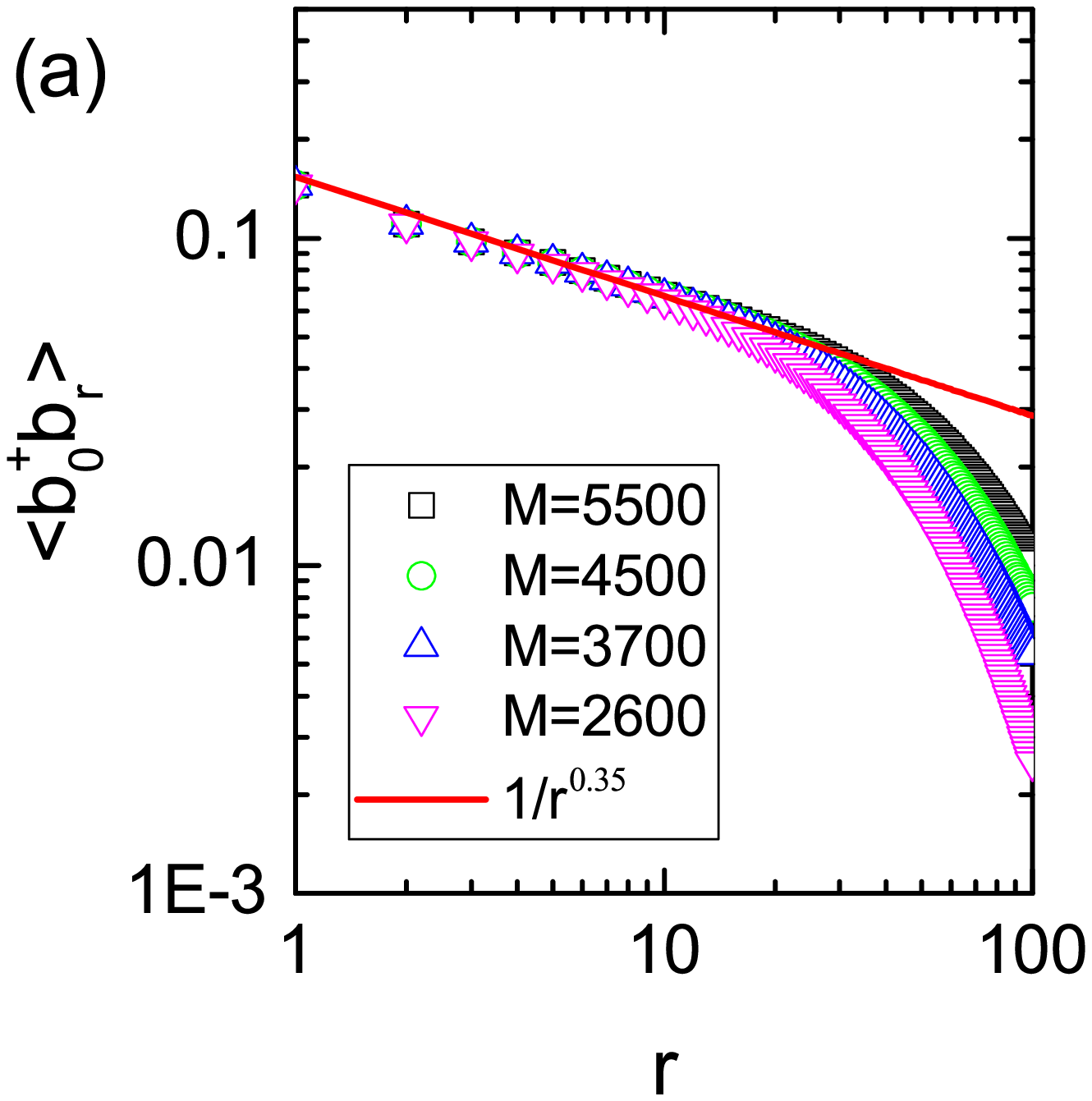}
  \includegraphics[height=1.85in,width=1.86in]{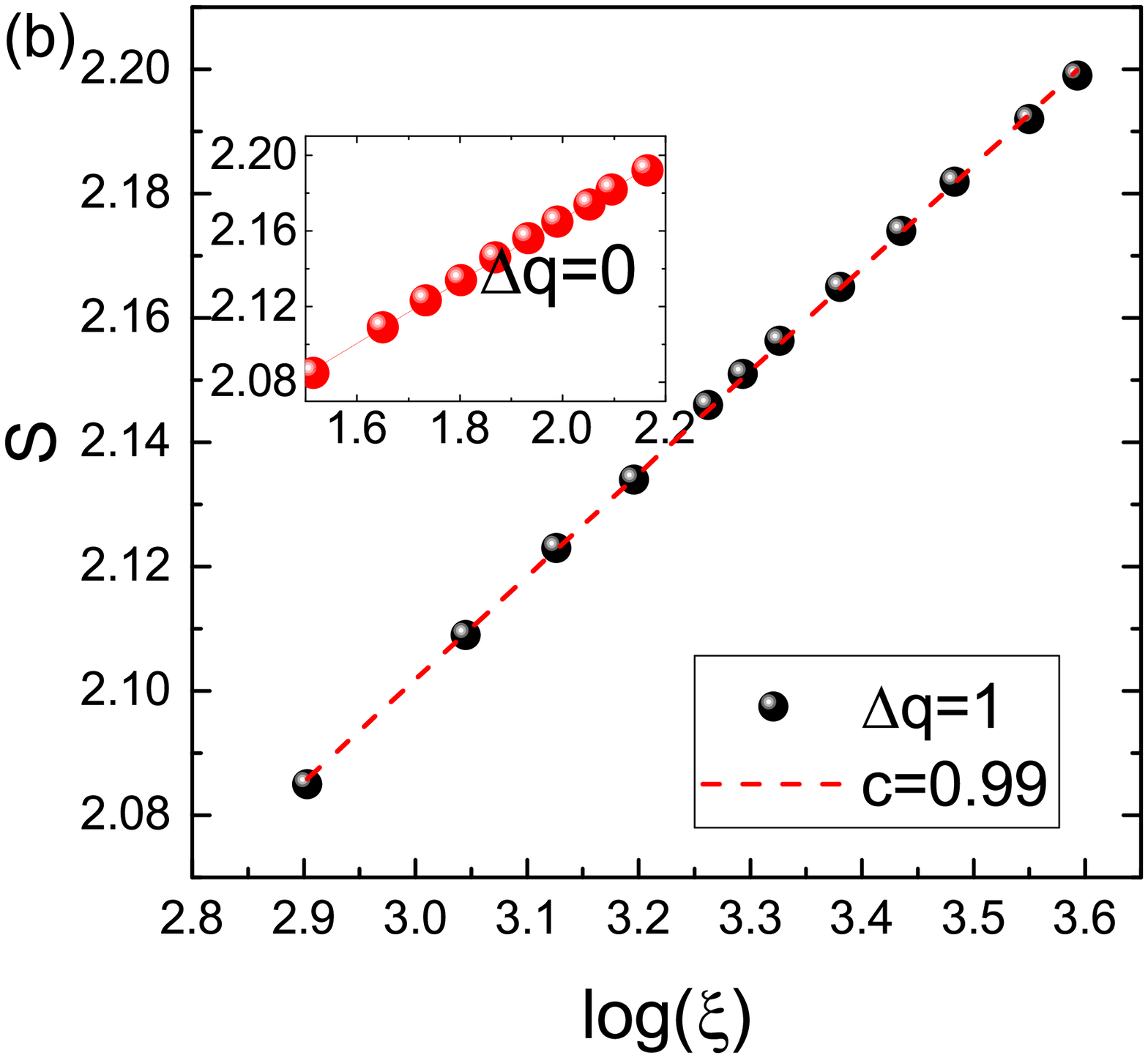}
  \caption{Numerical DMRG results near the critical point $\mu_c$ at $V=0$ on the infinite cylinder with width $L_y=8$. (a) The correlation function $\langle b^{\dag}_0 b_r\rangle$ and related power-law fitting (red line) with the increasing bond dimension $M$.
  Only the converged data that are independent of bond dimension $M$ are used for fitting.
  (b) Finite scaling of entanglement entropy vs correlation length for different charge sectors.
  The fitted dashed lines give the central charge $c\simeq1$ for both $\Delta q=0$ and $\Delta q=1$.}\label{fig:scaling}
\end{figure}

\section{Quantum criticality}\label{criticality}

Following the last section, we continue to analyze the phase transition nature from the BIQH phase
at weak potentials to the trivial (Mott) insulator at strong potentials with varying $\mu$. The possibility of continuous transitions between different quantum Hall states induced by periodic potentials was proposed in Refs.~\cite{Wen1993,Chen1993,Kol1993}, and has been intensively discussed regarding different Chern insulators in Ref.~\cite{Lee2018}.
Here, in our ED study, for two different lattice sizes $N_s=24,32$,
by tuning $\mu$ from weak to strong, no level crossing is observed between the ground state and low energy excited states.
In DMRG, we find that, both the first-order derivative of ground state energy per site $\frac{1}{N_s}\frac{\partial E(\mu)}{\partial \mu}$ and the density occupation difference $\Delta n$ between the majority and minority sites, evolve smoothly without
any singular behavior as a function of $\mu$, as indicated in Figs.~\ref{fig:corr}(a) and~\ref{fig:corr}(b).

To investigate the properties of the phase transition, we calculate the correlation lengths for different potentials $\mu$.
The correlation length is calculated using the transfer matrix defined in the infinite DMRG's algorithm~\cite{Ian2008}.
In Figs.~\ref{fig:corr}(c) and~\ref{fig:corr}(d), we plot the correlation length for different system sizes $L_y$,
which evolves continuously between the BIQH phase and trivial (Mott) insulator.
When the system is deep in either gapped phase (BIQH or trivial insulator),
the correlation functions of physical quantities are exponentially converged and the correlation lengths are very small.
In contrast, by approaching the critical point,
the correlation length reaches a broad peak.
As we increase the maximal bond dimension (the kept number of state) in the DMRG,
the correlation length tends to diverge at one unique peak signaling a critical point. For larger system widths such as $L_y=12$, more bond dimensions are demanded in order to ensure the convergence at the gapless critical point which is computationally difficult, and we focus our following discussions on a relatively small system width $L_y=8$ in our quasi-one-dimensional infinite cylinder. Nevertheless, for different system sizes, the smooth behavior of the correlation length supports a continuous phase transition.

In infinite DMRG, the correlation length $\xi(\Delta q)$ in the charge sector $\Delta q$ determines the largest correlation length of the operator-operator correlation function $\langle \widehat{O}^{\dag}_i\widehat{O}_j\rangle$ for the physical operator $\widehat{O}$ carrying the charge $\Delta q$.
Moreover, we also find that the correlation length $\xi$ in the charge sector $\Delta q=\pm1$
dominates over the other sectors, indicating that the slowest decaying channel is given by the correlation function $\langle b^\dagger_i b_j\rangle$ with $b^\dagger_i$ carrying the charge $\Delta q=1$.
The behavior of $\langle b^{\dag}_i b_j\rangle$ at the critical point is shown in Fig.~\ref{fig:scaling}(a).
It demonstrates an algebraically decaying behavior of single-particle off-diagonal long range correlations~\cite{Grover2013,Lu2014}  supporting a continuous transition.

\begin{figure}[t]
  \includegraphics[height=2.6in,width=3.0in]{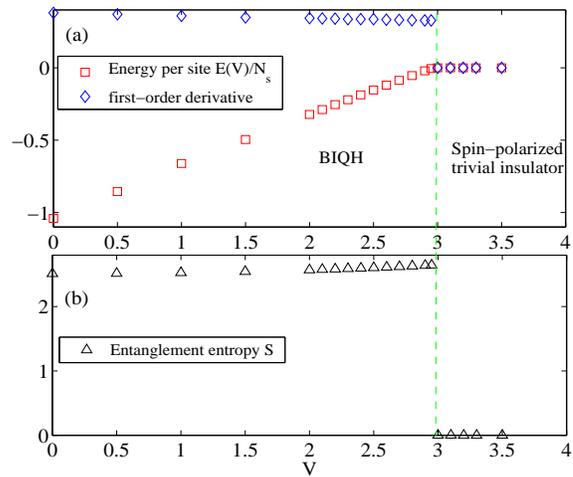}
  \caption{\label{int} The first-order phase transition between the BIQH and spin-polarized trivial insulator driven by interaction $V>0$ at $\mu=0$ on the infinite cylinder with width $L_y=8$:
  (a) the ground energy per site and its derivative vs $V$ and (b) entanglement entropy $S$.
  The green dashed line locates the transition point $V_c$.}
\end{figure}

Under the fermionic parton construction of bosons, this critical theory can be described by massless QED$_3$ with two flavors of fermions, which is dual to the easy-plane NCCP$^1$ model with two-component complex boson. In order to pin down the gapless nature of the critical point,
we calculate the finite scaling of entanglement entropy $S$ versus the correlation length.
We can learn more about the criticality by measuring the central charge $c$, which
can be thought of as the number of gapless modes. In conformal field theory, the central charge is extracted using the relation to the entanglement entropy~\cite{Calabrese2009}
\begin{align}
  S=\frac{c}{6}\log\xi+\text{const}.\label{ent}
\end{align}
For ordinary $N_f=2$ flavors of free massless Dirac fermions $\psi_{\alpha},\alpha=1,2$, the low-energy Lagrangian is
\begin{align}
  \mathcal{L}_{free}=\sum_{\alpha=1}^{N_f=2}\bar{\psi}_{\alpha}\gamma_{\mu}(-i\partial_{\mu})\psi_{\alpha},
\end{align}
where each fermion contributes a unit central charge, and the total central charge is simply given by $c=N_f$.
However, in deconfined criticality, the expected effective field theory around the critical point is an emergent QED$_3$ with two flavors of fermions $\psi_{\alpha},\alpha=1,2$ coupled to a gauge field, described by the Lagrangian
\begin{align}
  \mathcal{L}_{critical}=\sum_{\alpha=1}^{N_f=2}\bar{\psi}_{\alpha}[\gamma_{\mu}(-i\partial_{\mu}-\mathbf{a}_{\mu})]\psi_{\alpha}+\mathcal{L}[\mathbf{a}].\label{critical}
\end{align}
where $\mathbf{a}$ is a noncompact gauge field, and $\mathcal{L}[\mathbf{a}]$ is the Lagrangian for this gauge field. Different from $\mathcal{L}_{free}$, in Eq.~\ref{critical}, these two fermions are coupled to a dynamical U(1) gauge field $\mathbf{a}$. As a consequence, this gauge coupling acts as a finite mass gap, and would gap out the total charge mode~\cite{Geraedts2016},
which predicts an effective reduced central charge $c=N_f-1=1$ in Eq.~\ref{ent}. Figure~\ref{fig:scaling}(b) shows the scaling of $S$ near the critical point versus $\log\xi$ as the bond dimension $M$ increases, demonstrating a linear scaling law as Eq.~\ref{ent} with a linear slope $c\simeq1$. Therefore our numerical DMRG results rule out an ordinary Fermi liquid. Instead, they confirm the emergence of a gauge field in the critical point, which is the hallmark signature of the deconfined phase transition.

At last, we turn to consider the transition driven by intersublattice repulsive interaction $V$.
For strong interactions $V\gg1$, the system favors another topologically trivial insulator with suppressed particle entanglement between sublattice $A$ and sublattice $B$, namely a spin-polarized trivial (ferromagnetic) insulator with unequal particle numbers $N_{\uparrow}=\sum_j\langle n_j^{a}\rangle\neq N_{\downarrow}=\sum_j\langle n_j^{b}\rangle$ for the two-component spin structure.
In Figs.~\ref{int}(a) and~\ref{int}(b), we find the interaction $V$ driven transition is of first-order,
as evidenced by the discontinuous jump in energy derivative $\frac{1}{N_s}\frac{\partial E(V)}{\partial V}$
and a suddenly drop in entanglement entropy at the same transition point.

Physically, repulsive interaction $V$ favors the situation where particles totally occupy
in either sublattice A or sublattice B, leading to imbalanced particle fillings,
and would break the $U(1)\times U(1)$ symmetry down to global $U(1)$ symmetry,
in contrast to continuous transition driven by $\mu$ which preserves all symmetries with balanced particle fillings $N_{\uparrow}=N_{\downarrow}$.
The change of symmetry provides a natural understanding of the first-order transition [Fig.~\ref{fig:spt}(a)],
since a SPT phase is believed to be protected by underlying symmetries.
In this context, the phase transition driven by potential $\mu$ falls into the scenario of Fig.~\ref{fig:spt}(b),
where the protecting symmetry does not change across the transition and deconfined QCP can be realized.

\section{Summary and Discussions}\label{summary}

We have systematically investigated the phase transitions between a symmetry-protected topological (SPT) phase
and trivial insulators on a two-dimensional honeycomb lattice model that hosts a bosonic integer quantum Hall (BIQH) phase.
Using DMRG simulations, we find that the current model can realize both scenarios of Figs.~\ref{fig:spt}(a) and~\ref{fig:spt}(b)  under the interplay between interaction and periodic potential.
The main findings are as follows:
(1) we identify a direct continuous phase transition between BIQH and a featureless trivial (Mott) insulator;
and (2) we find evidences for the critical point as QED$_3$-Chern-Simons theory with two flavors of Dirac fermion coupled by
a gauge interaction,
thus supporting a realization of deconfined quantum criticality between a SPT phase and a trivial phase~\cite{Grover2013,Lu2014}.
By comparison, with increasing nearest-neighbor repulsions, there is a first-order transition from the BIQH phase to another trivial insulator, by breaking the protecting $U(1)\times U(1)$ symmetry.
In the future it would be interesting to explore the phase transition realizing the scenario of Fig.~\ref{fig:spt}(c),
e.g. the possibility of superfluid as an intermediate phase sandwiched between two SPT phases in a softcore bosonic model~\cite{Grover2013,Lu2014}.

\begin{acknowledgements}
T.S.Z. and W.Z. thank Zheng-Xin Liu, Yin-Chen He and Ying-Hai Wu for stimulating and helpful discussions.
This work is supported by start-up funding from Westlake University, and the NSFC under Grant No. 11974288 (T.S.Z. and W.Z.).
D.N.S. is also supported by the U.S. Department of Energy, Office of Basic Energy Sciences under Grant No. DE-FG02-06ER46305.
\end{acknowledgements}

\end{document}